# Linear_Accelerator_C+6_Ions_as_Injector_for_a_Synchrotron, Intended for Hadrons Therapy

*S. N. Dolya, K.A. Reshetnikova*

*Joint Institute for Nuclear Research, Joliot – Curie Street 6, Dubna, Russia, 141980*

**Abstract**

We consider acceleration of $C^{+6}$ ions by the field of a traveling-wave in a helical waveguide. The frequency of the accelerating RF field $f_0$ = 100 MHz, generator power $P_0$ = 2 MW. The initial energy of ions $E_{in}$ = 50 keV / nucleon, final energy $E_{fin}$ = 5 MeV / nucleon, accelerator length $L_{acc}$ = 10 m. A spiral with the initial and final radii of winding
$r_{0\,in}$ = 2.0 cm and $r_{0\,fin}$ = 1 cm is placed in a screen of diameter 2R = 100 mm. Ion focusing is provided by a solenoidal magnetic field with the intensity $B_0$ = 3.5 Tesla. With increasing the accelerator length up to L = 15 m, the final energy of the ions can be increased up to a value of $E_{fin}$ = 7 MeV / nucleon.

**Introduction**

For the treatment of cancer by hadrons therapy it is required to use $C^{+6}$ ions accelerated up to the energy 400 MeV / nucleon. These ions can be obtained from an isochronous or ion cyclotron. In the latter case, one requires an injector or linear accelerator of $C^{+6}$ ions (Z / A = ½) with an energy approximately equal to
$E_{inj}$ = 5 MeV / nucleon ($\beta_{inj}$ = 0.1, where $\beta$ = V / c is the speed expressed in terms of light speed in vacuum).

**1. Selection of basic parameters**

We assume the synchrotron perimeter P = 30 m. Then, for the one-turn injection the injected beam pulse duration $\tau$ = 1 μs. Now we assume that for treatment of one patient it is necessary to use a beam with intensity roughly equal to
$N_{C+6}$ = 3 * $10^9$ particles / s. Let the synchrotron cycle at the frequency F = 1 Hz. The initial energy of the accelerated ions can be of the order
$E_{in}$ = 50 keV / nucleon, ($\beta_{in}$ = 0.01) while no major problems occur with the high voltage power supply (U $_{e.s.}$ = 100 kV) and drift tube.

**2. Ion source**

Carbon ions and protons can be produced in required quantities and with the desired beam intensity by means of an ECR ion source. However, such a source is a complex and expensive device that poorly interfaces with the most suited for application in oncology synchrotrons operating in pulsed mode. A more appropriate source is the electron-beam ion source (EBIS).

**3. Linear accelerator**

Accelerators with an azimuthally symmetric wave have been traditionally used for ion acceleration. As is known [1], in this case one cannot provide both the radial and phase stability simultaneously because radial focusing under the conditions for auto phasing (radial focusing) requires presence of external field. In standing-wave accelerators quadruple focusing lenses were placed inside the drift tubes in the cavities, ion focusing by external longitudinal field being not in use at all then since the technology to provide constant high-intensity solenoid fields had not been yet developed.

In 1956 Vladimirskiy [2] proposed to introduce an azimuthally asymmetric component into the resonator gaps and implement focusing at the expense of the geometry of the accelerating field. Subsequently, the proposal was transformed into crystals with "fingers" and RFQ structure.

Such structures require high precision manufacturing and setting tolerances.

Through the development of technologies of superconducting solenoids it may become easier and cheaper to accelerate ions by an azimuthally symmetrical wave and perform their focusing by a longitudinal magnetic field.

In standing-wave resonators, the drift tubes (useless in terms of acceleration) occupy ¾ of the length of the accelerator, the entire field strength at the same time being concentrated in the gaps of the resonators operating in the prebreakdown regime. This makes no problem in a traveling-wave accelerator. The high-frequency pulse duration at acceleration by traveling wave can be short because there is no need in this case to excite the cavity.

**4. Accelerating structure based on a helical waveguide**

An accelerator-injector can be constructed on the basis of a spiral waveguide operating at the frequency $f_0 = 100$ MHz ($\lambda_0 = 3$ m), in which changes are made not only to the pitch but also to the radius of the frame onto which the spiral is wound to provide synchronism between particles and waves [3-4]. The frame is to have a conical shape with the initial radius of the spiral winding $r_{0in} = 2$ cm and final radius $r_{0fin} = 1$ cm, which can significantly increase the uniformity of the field along the helical waveguide.

When the power of the high-frequency generator $P_0 = 2$ MW and selected synchronous phase $\sin\varphi_s = 60^0$, the accelerator length turns short, approximately equal to $L_{acc} = 10$ meters, due to the acceleration in the traveling wave (the intensity of the wave field $E \approx 10$ kV / cm).

The spiral waveguide does not impose strict requirements for precision manufacturing and assembly structures. This is explained by the fact that a large number of turns of the spiral are involved in the formation of the wave field ($\lambda_{slow} = \beta\lambda_0 = 3\text{-}30$ cm, $h = 0.1\text{-}0.6$ cm, where $\beta$ is the particle velocity, $\lambda_0$ is the wavelength in the free space, $h$ is the step of the spiral winding) and, therefore, the allowance for the step of the spiral winding is not rigid.

Radial stability must be ensured by the external magnetic field $B_0 = 3.5\text{-}6$ T, created by a superconducting solenoid located on the outside of the spiral waveguide. The inner hole of the solenoid $D_i$, must be greater than the diameter of the outer conductor of the spiral waveguide $2R$, $D_i > 2R$ while the inner diameter of the solenoid must be of the order of $D_i = 10\text{-}15$ cm.

The screen located at a distance from the spiral and representing an outer conductor has virtually no effect on the electromagnetic properties of the spiral and does not cause any difficulties when manufacturing the superconducting solenoid.

The initial velocity of the particles of the carbon ion beam $\beta_{t0}$ obtained from an EBIS source lies in the range $\beta_{t0} = (\beta_{r0}^2 + \beta_{\varphi 0}^2)^{1/2} < 3 * 10^{-4}$. Evaluation of the phase space ellipse of the beam $\hat{E} = \pi * r_b * \beta_t / \beta_z$ yields the value $\hat{E} = \pi * 2 * (3 * 10^{-4}/0.02) = 30\pi *$ mm $*$ mrad, which can be achieved in the ion sources without any particular problems. In this case, the greater is the magnetic field of the $B_0$ coil which focuses the ions in a linear accelerator, the greater may be the initial radius of the particle beam.

**5. The structure of the field in a spiral waveguide placed in a screen**

Expressions for the fields in a spiral placed in an outer metallic cylinder (screen) can be written as [5], where index 1 refers to the region inside the spiral and index 2 to the region outside the spiral:

$$\begin{aligned}
E_{z1} &= E_0 I_0(k_1 r) \\
E_{r1} &= i(k_3/k_1) E_0 I_1(k_1 r) \\
H_{\varphi 1} &= i\varepsilon(k/k_1) E_0 I_1(k_1 r) \\
H_{z1} &= -i(k_1/\mu k) \text{tg}\Psi I_0(k_1 r_0) E_0 I_0(k_1 r)/I_1(k_1 r_0) \qquad (1) \\
E_{\varphi 1} &= -\text{tg}\Psi I_0(k_1 r_0) E_0 I_1(k_1 r)/I_1(k_1 r_0) \\
H_{r1} &= (k_3/\mu k) \text{tg}\Psi I_0(k_1 r_0) E_0 I_1(k_1 r)/I_1(k_1 r_0)
\end{aligned}$$

$$E_{z2} = I_0(k_1r_0)E_0[K_0(k_1r) - f_{01}I_0(k_1r)]/K_0(k_1r_0)d_0$$
$$E_{r2} = -i(k_3/k_1)I_0(k_1r_0)E_0[K_1(k_1r) + f_{01}I_1(k_1r)]/K_0(k_1r_0)d_0$$
$$H_{\varphi 2} = -i(k/k_1)I_0(k_1r_0)E_0[K_1(k_1r) + f_{01}I_1(k_1r)]/K_0(k_1r_0)d_0$$
$$H_{z2} = i(k_1/k)\text{tg}\Psi I_0(k_1r_0)[E_0K_0(k_1r) + f_1I_0(k_1r)]/K_1(k_1r_0)d_1 \quad (2)$$
$$E_{\varphi 2} = -\text{tg}\Psi I_0(k_1r_0)E_0[K_1(k_1r) - f_1I_1(k_1r)]/K_1(k_1r_0)d_1$$
$$H_{r2} = (k_3/\mu k)\text{tg}\Psi I_0(k_1r_0)E_0[K_1(k_1r) - f_1I_1(k_1r)]/K_1(k_1r_0)d_1,$$

where $k = \omega/c$, $k_3 = 2\pi/\beta\lambda_0$, $k_1 = (k_3^2 - k^2)^{1/2}$, $\omega = 2\pi f_0$, $d_0 = 1 - f_{01}I_0(k_1r_0)/K_0(k_1r_0)$, $d_1 = 1 - f_1I_1(k_1r_0)/K_1(k_1r_0)$, $f_{01} = K_0(k_1R)/I_0(k_1R)$, $f_1 = K_1(k_1R)/I_1(k_1R)$ and where, as in [5], the factor $e^{i(\omega t - k_3 z)}$ has been omitted.

The dispersion equation for a spiral placed in an outer metallic cylinder (screen) was obtained in [5]:

$$\text{ctg}^2\Psi = (k_1^2/k^2)I_0(k_1r_0)K_0(k_1r_0)/I_1(k_1r_0)K_1(k_1r_0)*\eta, \quad (3)$$

where

$$\eta = [1 - I_0(k_1r_0)K_0(k_1R)/I_0(k_1R)K_0(k_1r_0)]/[1 - I_1(k_1r_0)K_1(k_1R)/I_1(k_1R)K_1(k_1r_0)]. \quad (4)$$

An expression that connects power flux propagating in a shielded spiral with the electric-field strength on the axis for a spiral waveguide with a screen is as follows:

$$P = (c/8)*[kk_3/k_1^2]E_0^2 r_0^2 \{(I_1^2 - I_0I_2)*(1 + I_0K_1/I_1K_0\eta) +$$
$$+ 2[I_0^2 F_1/K_0^2 d_0^2 + I_1K_0F_2/I_0K_1\eta d_1^2]\}, \quad (5)$$

where $I_0$, $K_0$, $I_1$, $K_1$ are the values of the corresponding Bessel functions, given the argument equals $k_1r_0$,

$$F_1 = \int_{r_0}^{R} \check{Z}_1^2 r dr, \quad F_2 = \int_{r_0}^{R} \check{Z}_2^2 r dr, \quad (6)$$
$$\check{Z}_1 = K_1(k_1r) + f_{01}I_1(k_1r), \quad \check{Z}_2 = K_1(k_1r) - f_1I_1(k_1r). \quad (7)$$

As $R \to \infty$, the corresponding expressions tend to their limit values $f_{01} \to 0$, $f_1 \to 0$, $d_0 \to 1$, $d_1 \to 1$ and expressions for the fields in the area between the screen and spiral pass into expressions for the fields in a spiral in free space.

Expression (5) transforms into the corresponding formula [5] linking the power flow to the field strength on the axis for a spiral without a screen.

The effect of the screen consists in the reduction of the intensity of the external electric field on the helix axis, a corresponding reduction in the final energy of the particles (phase velocity of the wave) for a given length of the accelerator and a slight increase in the pitch [5].

**6. Linear accelerator. The longitudinal motion of ions**

A linear accelerator is to accelerate ions up to the speed $\beta_{\text{fin. lin.}} = 0.1$. In this respect, it is to start after pre-acceleration by the potential difference $U = 100$ kV with the speed $\beta_{\text{in. lin.}} = 0.01$, the initial energy being 50 keV / nucleon. With a speed increase up to tenfold, the energy will grow a hundredfold and the final energy will reach 5 MeV / nucleon.

Such an accelerator can be constructed on the basis of a spiral waveguide. The parameters of the high-frequency system and accelerator are about as follows: the acceleration frequency $f_0 = 100$ MHz, power $P_0 = 2$ MW, pulse duration $\tau_{\text{lin}} = 1$ microseconds, accelerator length $L_{\text{acc}} = 10$ meters.

Calculations of exact formulas for such parameters as the initial radius of the spiral $r_{0\text{ in.}} = 2$ cm, final radius of the spiral $r_{0\text{ fin.}} = 1$ cm, initial velocity $\beta_{\text{in. lin.}} = 0.01$, final speed $\beta_{\text{fin. lin.}} = 0.1$, radius of the screen $R = 5$ cm, acceleration rate $f_0 = 100$ MHz, power $P_0 = 2$ MW, focusing magnetic field $B_0 = 3$ T have showed that in this case the accelerator length $L_{\text{acc}} = 10$ meters. The desired pitch of the helix as a function of the acceleration length is illustrated in Fig. 1. The distribution of the electric field along the helix is presented in Fig. 2.

h, mm

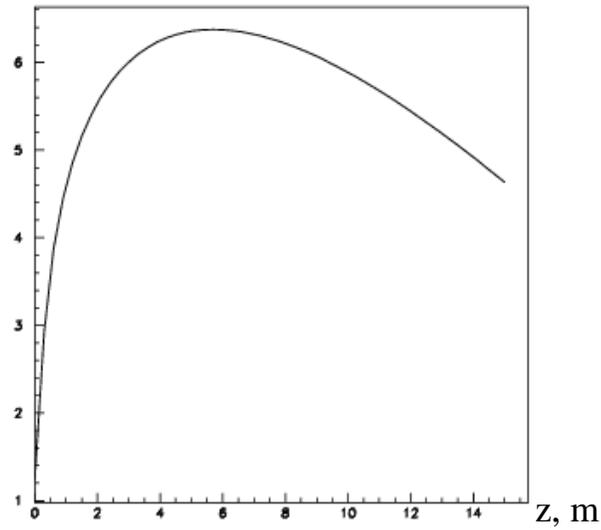

Fig. 1. The pitch of the helix winding as a function of the acceleration length.

The electric field remains roughly constant over the entire length of the accelerator.

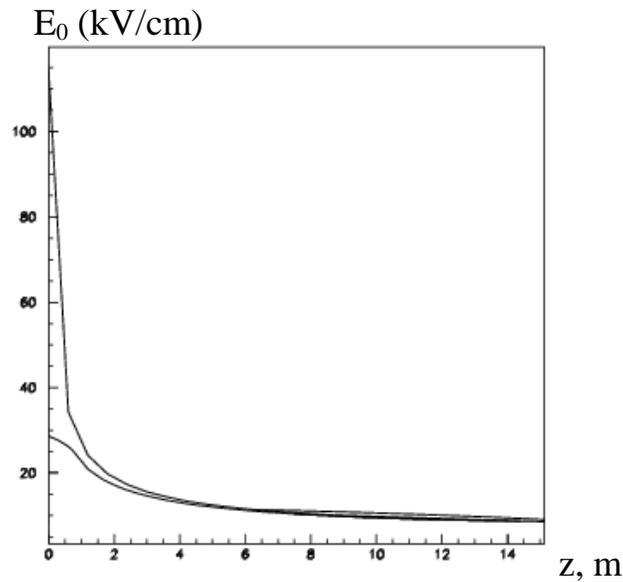

Fig. 2. The electric field strength at the helix (upper curve) and on its axis as a function of the accelerator length.

**7. Linear accelerator. The radial motion of ions**

The transverse Coulomb field $E_r$ can be estimated by the formula:

$$E_r = 2\pi Zen_i r_b = 500 \text{ V/cm}. \tag{8}$$

This field is added to the wave field $E_r = E_0 k_1 r/2 = E_0 * 2\pi r_b/\beta_i \lambda_0 = 3$ kV/cm, which is approximately 6 times greater for the chosen parameters than the Coulomb field on the surface of the beam. Fig. 3 shows the envelope of the beam with such initial parameters as $r_b < 0.2$ cm, $\beta_{r0} = +2*10^{-4}$, $\beta_\varphi = -2*10^{-4}$ in the magnetic field $B_0 = 3.5$ T.

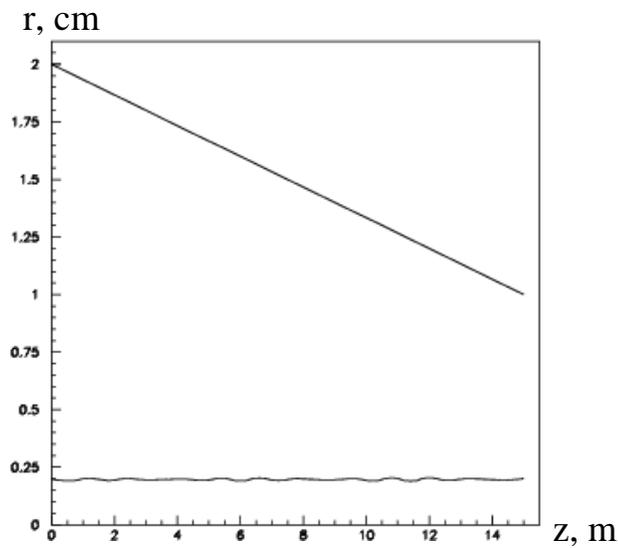

Fig. 3. The maximal radial deviation of the particle beam. The top line corresponds to the frame profile onto which the spiral is wound.

**8. Damping of waves in a spiral waveguide**

As a result of power damping in a spiral waveguide, the synchronous phase calculated for the case without damping will deviate from the synchronous phase obtained for acceleration of particles with account of attenuation.

Attenuation of waves propagating in the spiral is due to the ohmic losses of the currents flowing in the spiral. Power losses per turn are as follows:

$$\Delta P = \tfrac{1}{2} I_\varphi^2 * R. \qquad (9)$$

If the currents are expressed in Amperes and resistance in Ohms, the dimension $\Delta P$ is represented in Watts. Let us calculate the resistance of the coil: $R = 2\pi\rho * r_0 / (\pi\delta * d_{line})$. We assume that the coil is made of copper, then $\rho = 1.7 * 10^{-6}$ ohm*cm is the resistivity of copper, $r_0$ is the radius of the spiral turn, $r_{0in} = 2$ cm is the initial radius of the spiral winding, $r_{0fin} = 1$ cm is the final radius, $\delta = c / (2\pi\sigma\omega)^{1/2}$ is the depth of the skin layer; for copper $\sigma = 5.4 * 10^{17}$ s$^{-1}$, $\omega = 2\pi f_0$, $f_0 = 10^8$ MHz is the acceleration frequency. Let us assume that the thickness of the tape wound onto the spiral equals $d_{line} = 0.1$ cm

Instead of currents, it is convenient to substitute in (9) the magnetic fields associated with the currents:

$$I_\varphi^2 = (1.226)^{-2} * n^{-2} * H_z^2, \qquad (10)$$

where the dimension $H_z$, is presented in the Gaussian form, $n = 1 / h$ is the number of turns per centimeter of length, h is a pitch in the winding. Because one centimeter of the winding includes n turns wound onto it, one can finally obtain for $\Delta P = \tfrac{1}{2} I_\varphi^2 * R$ (W / cm):

$$\Delta P = (1.226)^{-2} * n^{-1} * H_z^2 * \rho * r_0 / (\delta * d_{line}). \qquad (11)$$

Let us make an estimate for the beginning of the spiral:

$n = 8/$cm, $H_z^2 = 4 * 10^4$ (Gs)$^2$, $\rho = 1.7 * 10^{-6}$ ohm*cm, $r_{0in} = 2$ cm, $\delta = 6.5 * 10^{-4}$ cm, $d_{line} = 10^{-1}$ cm, we obtain: $\Delta P = 200$ W / cm.

Now we introduce the quantity $2\alpha$, which is the effective length of power reduction:

$$\Delta P / P_0 = -2\alpha = 10^{-4} \text{ cm}^{-1}. \qquad (12)$$

The length $1 / \alpha$, equal to 200 meters, corresponds to a decrease of the field intensity by e times. Fig. 4 shows the magnetic field on the surface of the spiral coil as a function of the spiral length. Fig. 5 illustrates the dependence of the power propagating in the spiral on the length of the spiral coil.

$H_z$, Gs

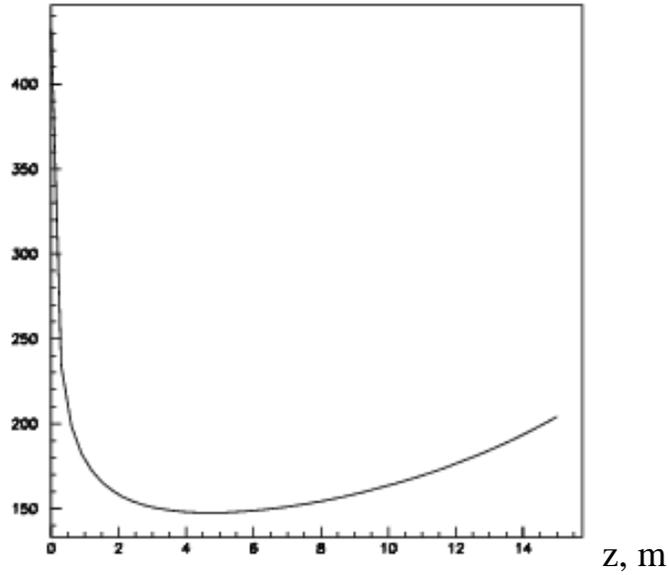

Fig.4. The magnetic field on the surface of the spiral.

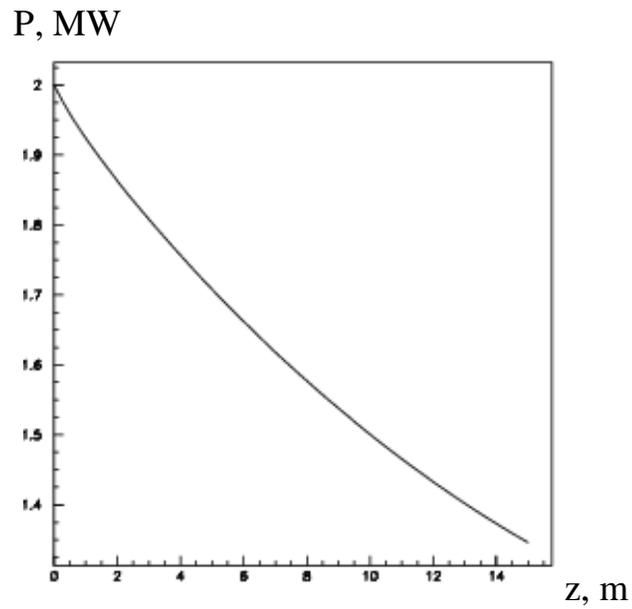

Fig. 5. RF power damping.

It is seen from Fig. 5 that damping does not play a decisive role in acceleration of ions in a helical waveguide. In Fig. 6. two curves show the dependence of the ion energy gain in the presence of damping (lower curve) and without damping. One can see that the difference in energy is not great at the accelerator length L = 15 meters.

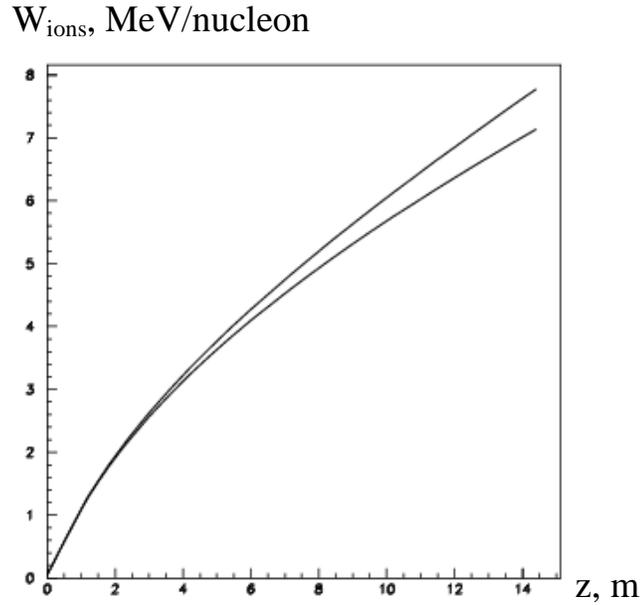

Fig. 6. The growth of ion energy as a function of the accelerator length. The initial ion energy $W_{ions}$ = 50 keV / nucleon; the lower curve is constructed with allowance for damping in the structure.

Table 1 shows the main parameters of the accelerator.

Table 1. The main parameters of the accelerator of $C^{+6}$ ions

| | |
|---|---|
| Frequency acceleration $f_0$, MHz | 100 |
| Pulse duration, μs | 1 |
| RF power generator, MW | 2 |
| Initial - final radius of the spiral, cm | 2 – 1 |
| Radius of the outer conductor (screen), cm | 5 |
| Initial energy of particles, keV / nucleon | 50 |
| Final particle energy, MeV / nucleon | 5 |
| Transverse velocities in the beam, $\beta_{t0} = (\beta^2_{r0} + \beta^2_{\varphi 0})^{1/2}$ | $< 3*10^{-4}$ |
| Initial radius of the ion beam $r_b$, cm | 0.2 |
| Focusing solenoidal magnetic field $B_0$, T | 3.5 |
| Accelerator length, m | 10 |

The magnetic field providing focusing grows in the proposed accelerator along with the initial radius of the beam.

In a synchrotron designed for hadron therapy the frequency and magnetic field tuning range has a value approximately equal to 10, from $\beta_{in}\gamma_{in} \approx 0.1 * 1 = 0.1$ to $\beta_{fin}\gamma_{fin} \approx 0.7 * 1.4 = 1$. In order to reduce the frequency or field tuning range, it is necessary to increase the injection energy. In the above case it can be achieved by increasing the acceleration length up to the value $L_{acc}$ = 15 meters, then $E_{ions}$ = 7 MeV / nucleon and the frequency tuning range is 1/0.12 = 8.33.